\documentclass[twocolumn]{autart}    

\usepackage{graphicx}          
\usepackage{afterpage}
\usepackage{epsfig} 
\usepackage{times} 
\usepackage{amsmath} 
\usepackage{amssymb}  
\usepackage{amsfonts}
\usepackage{mathptmx} 

\usepackage{tabularx}
\usepackage{epstopdf}
\usepackage{subfigure}
\usepackage{float}
\usepackage{epic,color}
\usepackage{mathrsfs}
\usepackage{dsfont,MnSymbol}
 \usepackage{soul,xcolor}

\usepackage{algorithm}
\usepackage{algorithmic}

\theoremstyle{definition}

\begin{document}
\setstcolor{red}
\begin{frontmatter}

\title{Regret Bounds for LQ Adaptive Control Under Database Attacks (Extended Version)}


\author[UIUC,CSL]{Jafar Abbaszadeh Chekan}\ead{jafar2@illinois.edu},    
\author[UIUC,CSL]{Cedric Langbort}\ead{langbort@illinois.edu},               

\address[UIUC]{Coordinated Science Lab., University of Illinois at Urbana Champaign}  
\address[CSL]{Department of Aerospace Engineering, University of Illinois at Urbana Champaign}             


\begin{abstract}                          
This paper is concerned with understanding and countering the effects of database attacks on a learning-based linear quadratic adaptive controller. This attack targets neither sensors nor actuators, but just poisons the learning algorithm and parameter estimator that is part of the regulation scheme. We focus on the adaptive optimal control algorithm introduced by Abbasi-Yadkori and Szepesvari and provide regret analysis in the presence of attacks as well as modifications that mitigate their effects. A core step of this algorithm is the self-regularized on-line least squares estimation, which determines a tight confidence set around the true parameters of the system with high probability. In the absence of malicious data injection, this set provides an appropriate estimate of parameters for the aim of control design. However, in the presence of attack, this confidence set is not reliable anymore. Hence, we first tackle the question of how to adjust the confidence set so that it can compensate for the effect of the poisonous data. Then, we quantify the deleterious effect of this type of attack on the optimality of control policy by bounding regret of the closed-loop system under attack.
\end{abstract}

\end{frontmatter}

\section{Introduction}

Due to their often distributed nature and reliance on numerous and disparate units (e.g. sensors, actuators, computing units and databases) cyberphysical systems are particularly susceptible to attacks aimed at disrupting their operations or bringing them down. A particular kind of attack that has received significant attention in the past few years is fake data injection (FDI) attacks, whereby some signals circulating in the feedback loop are strategically modified by an adversarial entity \cite{mo2010false,hendrickx2014efficient}. Examples include the notorious Stuxnet attack \cite{langner2011stuxnet} on supervisory control as well as proof of concept attacks on water SCADA systems \cite{amin2010stealthy}.

One reason why FDIs have been considered with such particular attention is that they represent an \textit{indirect} attack on a controlled system (as opposed to, e.g., breaking down the infrastructure or introducing adversarial process noise): the fact that  some data is corrupted only matters to the extent that it affects closed-loop performance, be it because it prevents correct state estimation or the computation of the correct control signal. What the data is used for determines the impact of the FDI and, accordingly, the necessary defense mechanisms.

In control schemes where data is used for purposes other than state estimation (e.g., in data driven approaches \cite{berberich2020robust} or adaptive/learning-based controllers, where it is also used to identify the plant in real time) FDI attacks may thus have a different effect on a controller's ability to regulate a system of interest. Indeed, the effects of data attacks on a learning-based algorithm operating in the loop are not intuitively clear. On the one hand, one may reason that the attack will be quite damaging, since false data not only triggers mis-estimation of the system's model, but this mis-estimation may itself result in
the computation and injection of an incorrect control input
signal, further driving the system's state away from its desired
optimal value. On the other hand, one could argue that
because the controller constantly adapts itself and receives
new measurements, it might be able to correct the effects
of an attack if it is limited in space and time. 

In order to rigorously address this question in a restricted yet relevant setting, we study a specific kind of data attack on a class of learning-based algorithms for the linear quadratic regulator (LQR) problem. We show that, while closed-loop performance can degrade rapidly in the presence of attack (both in terms of regret and of the algorithm's ability to maintain an appropriate estimate of the system's parameters), it is possible to modify the algorithm's online estimation procedure to recover a guaranteed sub-linear regret under attacks. The restricted setting we investigate has the following properties:

\begin{itemize}
\item We only consider model-based online learning algorithms which, unlike model-free learning-based techniques (like, e.g, \cite{abbasi2019model}, \cite{arora2018towards}, \cite{fazel2018global}, \cite{tu2018least}), assume a particular model structure and continuously identify the parameters of this model at the same time as they produce a control input (which constitutes the ``online learning" element). Such algorithms rely on the paradigm of certainty equivalence familiar in adaptive control, and have been shown to be more data efficient than model-free alternatives \cite{clavera2018model}, even provably achieving minimal sample complexity in the LQR setup \cite{tu2019gap}. Work on this class of algorithms started with \cite{campi1998adaptive} which introduced a cost-biased parameter estimator and relied on the principle of ``Optimism in the Face of Uncertainty'' (OFU) to provably achieve asymptotically optimal closed-loop cost for linear quadratic Gaussian systems. Later, Abbsi-Yadkori and Szepesvari went beyond these asymptotic guarantees and proposed another online-learning based algorithm with a guaranteed regret bound of $O(\sqrt{T})$ in $T$ rounds \cite{abbasi2011regret}. More recently, Ibrahimi et al. \cite{ibrahimi2012efficient} followed similar lines and proposed an algorithm
that achieves $O(p\sqrt{T})$ regret bound for plant of state space
dimension $p$, while the authors of \cite{faradonbeh2020optimism} managed to retain a $O(\sqrt{T})$ regret bound for their OFU-based learning algorithm under milder assumptions. Cohen et al. \cite{pmlr-v97-cohen19b}, while keeping loyalty to the main idea of the algorithm in \cite{abbasi2011regret}, proposed a computationally efficient algorithm by formulating the LQR
control problem as a semi-definite programming problem
and, as such, resolved the open question of the literature
regarding efficient computation of a proper certainty equivalent
controller at each time step. Furthermore, in \cite{lale2020explore}
the authors adds an additional step of ``explore more'' to
the algorithm in \cite{abbasi2011regret} which gives better general regret bound
thanks to obtaining an initial stabilizing control.

It is the existence of tight rigorous regret bounds for OFU online-learning based algorithms that makes them particularly well-suited for quantitatively evaluating the effect of data attacks. 

\item We consider a specific type of FDI, called ``database attack'', in which plant measurements are tampered with only after they are stored for later processing, as opposed to being modified directly at the sensors' level. Such storage occurs, e.g., in the context of cloud robotics \cite{waibel2011roboearth} or cloud-based control \cite{alexandru2020secure}, where data is uploaded in real-time to a remote cloud repository before being accessed as needed by a controller co-located with the plant, or where the controller is itself delocalized in the cloud. If illicit access to the repository can be gained (a feat that may be easier to achieve for an attacker than targeting a physical sensor e.g., via an SQL injection or simply by getting a hold of authorized credentials) \cite{linkref},   the stored data can be modified, as was done, e.g., in a recent example attack on a container ship aimed at redirecting its shipments \cite{linkref}. In the case of the learning-based algorithms discussed above, data is stored (or at least put in memory) to set up the estimation problem at every time step, and a database attack can thus affect the learned value of the parameters.  While we should expect an attacker to rely on all available vulnerabilities and to conduct more complex data attacks in practice, focussing solely on database attacks (as opposed to more general FDIs)  is valuable from a theoretical viewpoint, as this class separates the effects of this mis-learning of system parameters on the closed-loop regret from those due to state mis-estimation or control input mis-computation.
\end{itemize}

The remainder of the paper is organized as follows. Section
\ref{sec:prelim} presents the class of online-learning based algorithms of interest, and reviews preliminaries and background  as presented in \cite{abbasi2011regret}, in particular the central notion of confidence set.  The attack model is detailed in Section \ref{sec:ProbFor}, which also illustrates the loss of performance experienced by this controller. In Section \ref{section4}, we show how to modify the confidence set so as to recover a satisfactory regret bound in the presence of the attack. Finally, Section
\ref{sec:simulation} illustrates the paper's key contributions by providing simulation results.

\section{A review of online-learning based LQR control}
\label{sec:prelim}

In this section a background review of online-learning based adaptive control in the LQR context is presented. We summarize the central concepts and results of \cite{abbasi2011online} and \cite{abbasi2011regret}, which serve as the main building blocks for the design of a database attack-resilient controller in Section 4.

Consider the following linear time invariant dynamics and the associated cost functional given by:
\begin{subequations}
	\begin{align}
		x_{t+1} &=A_{*}x_{t} + B_{*}u_{t}+\omega_{t+1}\label{eq:dyn}\\
		c_{t} &=x_{t}^T Qx_{t} + u_{t}^T Ru_{t}\label{eq:obs}
	\end{align}
\end{subequations}
where the plant and input matrices $A_{*}\in  \mathbb{R}^{n \times n}$ and  $B_{*}\in  \mathbb{R}^{n \times m}$ are initially unknown and have to be learned. {$Q\in  \mathbb{R}^{n \times n} $ and $R\in  \mathbb{R}^{m \times m}$ represent known and positive definite and semi-definite matrices respectively} and $\omega_{t+1}$ is noise signal. 
The associated average expected cost based on the past observations is written as:
\begin{align}
	J(\pi) =\lim_{T\to\infty} \frac{1}{T}  \sum _{t=1}^{T} \mathbb {E} [c_{t}] \label{eq:Ave_cost} 
\end{align}
where $u_0$, $u_1$,..., $u_{T-1}$ are chosen based on the policy $\pi$ starting from $x_0$. The regret of this strategy is defined as 
\begin{align}
	R_T =\sum _{t=1}^{T} (\mathbb {E} [c_{t}]-J^*)\label{eq:Reg} 
\end{align}
where $J^*$ is the cost of optimal control strategy computed with knowledge of the matrices $A_{*}$ and $B_{*}$. 
$R_T$ is a measure of how much the lack of insight into the model affects performance. By defining,
\begin{align}
	\Theta _{*}^T =(A_{*},\,\,\,     B_{*})\label{eq:def_theta} 
\end{align}
the system transitions dynamics can be rewritten as:
\begin{align}
	x _{t+1} =\Theta _{*}^T z_{t}+\omega_{t+1}, \quad z_t=\begin{pmatrix} x_t \\ u_t \end{pmatrix} \label{eq:dynam_by_theta} 
\end{align}
In the analysis of our setting we will make the following core assumptions,

\begin{assum} There exists a known set $S\subset S _{0}\cap S _{1}$, to which $\Theta_*$ belongs where
\begin{align*}
	&S _{0} =\{\Theta \in \mathbb{R}^{(n+m) \times n}  \mid trace(\Theta^T\Theta)\leq \vartheta^2  \} \; \;  \text{for some $\vartheta >0$,} \\
	& S _{1} =\{\Theta=(A,B) \in \mathbb{R}^{(n+m) \times n}  \mid  \text{$(A,B)$ is controllable,}\\ 
	&\quad\text{$(A,M)$ is observable, where $Q=M^TM$}  \}.
\end{align*}
\end{assum}

\begin{assum}
There exists a filtration $\mathcal{F}_{t}$ such that

$(2.1)$ $z_{t}$ and $x_{t}$ are $\mathcal{F}_{t}$-measurable.

$(2.2)$ for any $t \geq 0$,
 \begin{align*}
	\mathbb{E}[x_{t+1}|\;\mathcal{F}_{t}]=\Theta_{*}^Tz_{t}
\end{align*}
$(2.3)$
 $\mathbb{E}[\omega_{t+1}\omega_{t+1}^T|\; \mathcal{F}_{t}]=I_{n}$;
 
$(2.4)$ $\omega_{t}$ are component-wise sub-Gaussian i.e. there exists $L>0$ such that for any $\gamma \in R$ and $j=1,2,...,n$
\begin{align*}
	\mathbb{E}[e^{\gamma(\omega_{t+1})_j}|\;\mathcal{F}_{t}]\leq e^{\gamma^2L^2/2}.
\end{align*}
\end{assum}

{Assumption 2 on the process noise is a standard assumption in both the controls (see \cite{abbasi2011regret}, \cite{cohen2019learning}, and \cite{lale2022reinforcement}) and bandit (see \cite{abbasi2011improved}, \cite{agrawal2013thompson}, and \cite{vakili2021optimal}) literatures.}

Using the self-normalized process, the least square estimation error up to time $t$, $e(\Theta)$ can be obtained as:
\begin{align}
	e(\Theta)&= \lambda  \operatorname{Tr}(\Theta^T\Theta)+\sum _{s=0}^{t-1} \operatorname{Tr} \big((x_{s+1}-\Theta^Tz_{s})(x_{s+1}-\Theta^Tz_{s})^T)\big).  \label{eq:LSE_op}
\end{align}
where $\lambda$ is a regularization parameter.
This yields the $l^{2}$-regularized least square estimate:
\begin{align}
	\hat{\Theta}_{t} &=\operatorname*{argmin}_\Theta e(\Theta)=(Z^T_tZ_t+\lambda I)^{-1}Z^T_tX_t \label{eq:LSE_Sol}
\end{align}

where $Z_t$ and $X_t$ are matrices whose rows are $z_{0}^T,...,z_{t-1}^T$ and $x_{1}^T,...,x_{t}^T$, respectively.

Defining covariance matrix $V_{t}$ as follows:
\begin{align*}
	V_{t}=\lambda I + \sum_{s=0}^{t-1} z_{s}z_{s}^T=\lambda I +Z^T_tZ_t,
\end{align*}
\cite{abbasi2011regret} shows that with probability at least $(1-\delta)$, where $0<\delta<1$, the true parameters of system $\Theta_{*}$ belong to the confidence set defined by: 
\begin{align}
	C_{t}(\delta) =\{\Theta \in \mathbb{R}^{n \times (n+m)}  \mid \operatorname{Tr}((\hat{\Theta}_{t}-\Theta)^TV_{t}(\hat{\Theta}_{t}-\Theta))\leq \beta_{t}(\delta)\} \label{eq:Conf_Set_unaata} 
\end{align}
where 
\begin{align}
	\beta(\delta) =\bigg(nL\sqrt{2\log(\frac{\det(V_{t})^{1/2}\det(\lambda I)^{-1/2}}{\delta}})+\lambda^{1/2}\vartheta \bigg)^{2} \label{eq:Conf_set_radius_unAtt}. 
\end{align}

By the controllability and observability assumptions (assumption 1) there exists a unique positive definite solution $P(\Theta)$ to the algebraic Riccati equation (ARE):

\begin{align*}
	P(\Theta)= Q+A^TP(\Theta)A-A^TP(\Theta)B(B^TP(\Theta)B+R)^{-1}B^TP(\Theta)A
\end{align*}
for all $(A,B) \in S$.
Under this assumption the linear optimal control law $u(t)=K(\Theta)x(t)$ where
\begin{align*}
	K(\Theta)= -(B^TP(\Theta)B+R)^{-1}B^TP(\Theta)A.
\end{align*}
is stabilizing, i.e.$ \left\lVert (A+BK(\Theta))\right\rVert <1$ and the average cost of control law with $\Theta=\Theta_{*}$ is the optimal average cost $J(\Theta_{*})=trace(P(\Theta_{*}))$.

In addition, boundedness of $S$ results in boundedness of $P(\Theta)$ and $K(\Theta)$ with constants $D$ and $C$ respectively:
\begin{align*}
	D&=Sup \{\left\lVert P(\Theta)\right\rVert \mid\Theta \in S \},\\
	C&=Sup_{\Theta \in S}||K(\Theta)||<\infty.
\end{align*}
After finding high-probability confidence sets for the unknown parameter at time $t$, the core step of the algorithm proposed in \cite{abbasi2011regret} is implementing the Optimism in the Face of Uncertainty (OFU) principle. At any time $t$, we choose a parameter  $\tilde{\Theta}_t \in S\cap C_{t}(\delta)$ such that:
\begin{align}
	J(\tilde{\Theta}_t )\leq \inf\limits_{\Theta \in C_t(\delta)\cap S }J(\Theta)+\frac{1}{\sqrt{t}}. \label{eq:nonconvexOpt} 
\end{align}
Then, by using the chosen parameters as if they were the true parameters, a stabilizing controller is designed by solving the ARE. As can be seen in the regret bound analysis of \cite{abbasi2011regret}, recurrent switches in policy may worsen the performance, so a criterion is needed to prevent frequent policy switches. As such, at each time step $t$ the algorithm checks the condition $det(V_{t})>2det(V_{\tau})$, where $\tau$ is last policy update time, to determine whether updates to the control policy are needed. Algorithm 1, adopted from \cite{abbasi2011regret}, provides the detailed procedure.

\begin{algorithm} 
	\caption{Adaptive Algorithm for LQ control problem} \label{alg:ACLQ}
	\begin{algorithmic}[1]
   	    \STATE \textbf{Inputs:} $T,$$\,s>0,$$\,\delta>0,$$\,Q\,, L,\, \lambda>0$
		\STATE set $V_0=\lambda I$ and $\hat{\Theta}=0$
		\STATE $\tilde{\Theta}_0=argmin_{\Theta \in C_0(\delta)\cap S}\,\, J(\Theta)$
		\FOR {$t = 0,1,2,...$}
		\IF {$\det(V_t)> 2 \det(V_{\tau})$ or $t=0$} 
		\STATE Calculate $\hat{\Theta}_t$ by (\ref{eq:LSE_Sol}) and set $\tau=t$
		
		\STATE Find $\tilde{\Theta}_t$ such that $J(\tilde{\Theta}_t)\leq \inf_{\Theta \in C_t(\delta)\cap S} J(\Theta)+\frac{1}{\sqrt{t}}. $
		\ELSE
		\STATE $\tilde{\Theta}_t=\tilde{\Theta}_{t-1}$
		\ENDIF 
		\STATE For the parameter $\tilde{\Theta}_t$ solve ARE and calculate $u_t=K(\tilde{\Theta}_t)x_t$
		\STATE Apply the control and observe new state $x_{t+1}$.
		\STATE Save $(z_t,x_{t+1}) $ into dataset
		\STATE $V_{t+1}=V_t+z_tz^T_t$
		\ENDFOR
	\end{algorithmic}
\end{algorithm}

The policy explicited in Algorithm 1 keeps the states of the underlying system bounded with probability $1-\delta$ which is defined as the "good event" $F_{t}$:
\begin{align}
	F_{t}=\{\omega \in\Omega  \mid \forall s \leq t, \left\lVert x_{s}\right\rVert \leq \alpha_{t} \}.\label{eq:GoodEven_state_unat} 
\end{align}
A second "good event" is associated with the confidence set defined as:
\begin{align}
	E_{t}=\{\omega \in\Omega  \mid \forall s \leq t, \Theta_{*} \in C_{s}(\delta/4)  \}\label{eq:GoodEven_set_unat} 
\end{align}
where both good events are defined in probability space $\Omega$ in which the noise $\omega_{t+1}$ takes its values and $\alpha_{t}$ has been explicited in \cite{abbasi2011regret}. 

Finally, if we let $E=E_{T}$ and $F=F_{T}$, then it is proven in \cite{abbasi2011regret} that intersection of $E$ and $F$ holds with high probability i.e. $P(E\cap F)\geq 1-\delta/2$.

By an appropriate decomposition of regret on the event $E\cap F$, it is also shown that:
\begin{align}
	R_T \leq R_1-R_2-R_3+2\sqrt{T}\label{eq:regret_unat}   
\end{align}
where 
\begin{align}
	R_1=\sum _{t=0}^{T} (x_{t}^T P(\tilde{\Theta}_{t})x_{t}-\mathbb{E}[x_{t+1}^T P(\tilde{\Theta}_{t+1})x_{t+1}|\;\mathcal{F}_{t}])\label{eq:R1}  
\end{align}
\begin{align}
	R_2=\sum _{t=0}^{T} \mathbb{E}[x_{t+1}^T(P(\tilde{\Theta}_t)-P(\tilde{\Theta}_{t+1}))x_{t+1}|\;\mathcal{F}_{t}]\label{eq:R2}  
\end{align}
and
\begin{align}
	\nonumber R_3 &=\sum _{t=0}^{T} \big((\tilde{A}_tx_{t}+\tilde{B}_tu_{t})^T P(\tilde{\Theta}_t)(\tilde{A}_tx_{t}+\tilde{B}_tu_{t})\\
	&\quad-(A_{*}x_{t}+B_{*}u_{t})^T P(\tilde{\Theta}_{t+1})(A_{*}x_{t}+B_{*}u_{t})\big)\label{eq:R3}
\end{align}
By bounding the above terms separately, for any $T$ it is shown that with probability at least $1-\delta$, the imposed regret is:
\begin{align}
	\nonumber R &\leq 2D W^2 \sqrt{2T \log \frac{8}{\delta}}+n\sqrt{B^\prime_{\delta}}+ 2D X^2_{T}(n+m) \\
	\nonumber &\quad\log_2 (1+2T/\lambda(X^2_{T}(1+C^2)))  + \frac{8}{\sqrt{\lambda}}((1+C^2)X^2_{T})\\
	&\quad sD\sqrt{\beta_T(\delta/4)}\Bigg(\log \frac{det(V_t)}{det(\lambda I)}\Bigg)^{1/2}\sqrt{T}\label{eq:regret_unaatacked}
\end{align}	
where $\max_{1 \leq s \leq T}||x_s||\leq X_T$, $W=Ln\sqrt{2n\log(8n/\delta)}$ (from sub-Gaussianity assumption) and definition of $B^\prime_{\delta}$ is provided in \cite{abbasi2011regret}. As can be seen, the regret is $O(\sqrt T)$ up to time $T$. 

\section{Database Attacks and Performance Degradation}
\label{sec:ProbFor}
We are interested in the situation depicted in Figure 1, where a controller designed based on Algorithm 1 operates in an adversarial environment.

As can be seen in Figure 1, at each time instant $t$, the plant's state $x^a_t$ is sent both to the controller block (to compute and implement control input $u^a_t = K(\tilde{\Theta}_t) x_t^a$), and to a database, where it is used to generate the covariance matrices $V_s, s\geq t$ which are involved in the estimation of $\Theta$ from then on. Note the ``a'' superscript will be used from now on to denote the fact that we are considering state and control input of a system under attack (as opposed to the discussion of Section \ref{sec:prelim}). A similar notation will be used for extended state $z_t^a$ as well, which contains both state and control input of the attacked system at $t$.

\begin{figure}[thpb]
	\centering
	\vspace{5pt}
	\includegraphics[scale=.4]{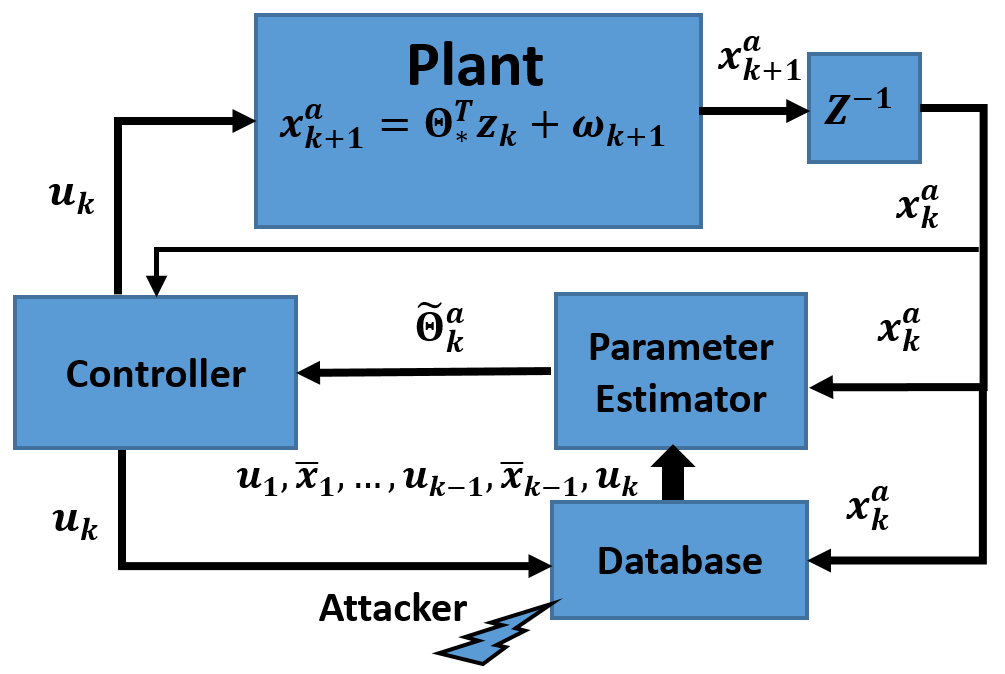}

	\caption{Diagram of a closed-loop system with the learning block, attacked in the loop. Notice that the database transmits corrupted past values of state ($\bar{x}\;\; \forall s=1,...,k-1$) to Parameter Estimator}.
	\label{figurelabel}
\end{figure}

In the most general scenario, an attacker would likely disrupt all elements of the feedback loop, from poisoning the data stored in the database to blocking the transmission of the control input to modifying the estimation algorithm, and it would be necessary for the control schemes to be robust, in some sense, to all these manipulations \cite{teixeira2012attack}. However, since the role played by stored data is unique to the control scheme presented in Section \ref{sec:prelim}, we focus solely on data poisoning attacks on the database in this work. We will henceforth refer to this attack model as a "database attack".

Accordingly, we assume that, at time $t$, (1) the controller receives the correct value $x^a_t$ and, (2) this value is also correctly stored in the database at time $t$. 
However, (3) the attacker replaces the true values $x^a_1,..., x^a_{t-1}$ of the state by $\bar{x}_1, ..., \bar{x}_{t-1}$. We place no restriction on the attacker's strategy beyond the assumption that there exists $\Lambda$, known to the controller, such that
 \begin{align}
	\Lambda \geq max_{ 1\leq s  \leq T} || \bar{x}_s - x^a_{s} || \label{eq:attacked_st}
\end{align}	
 This means, in particular, that the attacker can have full knowledge of the plant model (even though it is unknown to the controller) and exploit this knowledge strategically, and that the attack signal $\eta_s := \bar{x}_s - x^a_{s}$ can satisfy any arbitrary dynamics or even be stochastic, as long as it remains bounded. We note that the bound $\Lambda$ can be loose with respect to the actual maximum of $\eta_s$ over $[1,T]$, which means that the controller need not know the exact extent of the attack, and can be conservative in planning its response. While an attacker's ability to inject arbitrary (i.e., not a priori bounded) signals has sometimes been considered as a hallmark of false data injection attacks in the literature \cite{fawzi2014secure,mao2022computational}, models of such attacks as unknown but bounded signals have also been proposed \cite{degue2018interval}. Assuming that $\eta$ is unknown but bounded makes sense in the present context, since too large deviations may easily be spotted by appearing to be obviously inconsistent with the value $x^a_t$ available to the controller at time $t$.
 
Item (2) above is meant to capture the fact that data is checked as it is deposited in the database but not later (in such a way that a modification of $x^a_t$ at time $t$ could simply be noticed by comparing the stored value with that received by the controller, thus rendering such an attack useless). Other similar restrictions on the values that the attacker is permitted to modify in the database at time $t$ (e.g., $x_i^a$ for $i \leq t-k$) could also be considered without affecting much of our analysis, as a way to account for longer memory at the controller. We would argue, however, that once an implementation that separates "controller"-block and "database" has been chosen, it is natural to expect that this memory is finite and short (since, presumably, the reason why a database is used is that the controller has reduced storage space), thus making it possible for the attacker to modify at least \textit{some} stored past values of the state at each time $t$ without being noticed.
\hfill 

Under this attack model, the dynamics of the stored poisoned data is given by
\begin{align}
	\bar{X}_t &=Z^{a}_t \Theta _{*}+W_t+H_t. \label{eq:dyPttt}
\end{align}
where $\bar{X}_t$ and $Z^a_t$ are matrices whose rows are $\bar{x}^T_1, ...,\bar{x}^T_t$ and $z^{aT}_0, ...,z^{aT}_{t-1}$ respectively. Also, $H_t$ is a matrix whose rows are $\eta_1^T,...,\eta_{t-1}^T,0^T$ and $W_t$ is a matrix with rows, $\omega_1^T,...,\omega_t^T$. 

Then, for history of data $\bar{Z}_t$ given by dataset we have:
\begin{align}
	\bar{Z}_t &=Z^{a}_t+Y_t \label{eq:BarZi_Z^aRel}
\end{align}
where $Y_t$ is a matrix with rows $\zeta^T_1,...,\zeta^T_t$ defined as follows:
\begin{align*}
	\quad \zeta_t=\begin{pmatrix} \eta_{t+1} \\ 0_{(m\times 1)} \end{pmatrix}.
\end{align*} 
Having $\bar{Z_t}$ as the matrix whose rows are $\bar{z}_{1}^T,...,\bar{z}_{t-1}^T$, similar to (\ref{eq:LSE_op},\ref{eq:LSE_Sol}), the normalized least square error can be written as:
\begin{subequations}
	\begin{align}
		e(\Theta)&= \lambda  \operatorname{Tr}(\Theta^T\Theta)+\sum _{s=0}^{t-1} \operatorname{Tr}\big((\bar{x}_{s+1}-\Theta^T\bar{z}_{s})(\bar{x}_{s+1}
		-\Theta^T \bar{z}_{s})^T\big)  \label{eq:LSE_attck}\\
		\hat{\Theta}_{t}^a &=\operatorname*{argmin}_\Theta e(\Theta)=(\bar{Z_t}^T\bar{Z_t}+\lambda I)^{-1}\bar{Z_t}^T\bar{X_t}. \label{eq:LSE_sol_att}
	\end{align}
\end{subequations}
As a result of matrix $Z^a_t$ being replaced by $\bar{Z}_t$ in (\ref{eq:LSE_sol_att}), the system's parameters are misestimated, which in turn may result in poor performance of the controller or even destabilization of the closed loop if Algorithm 1 is just implemented as-is. It is also possible for the algorithm to completely fail if the set S fails to intersect $C_t$ at some time $t$. 

\begin{figure}
	\centering
	\vspace{5pt}
	\begin{subfigure}
		\centering
		\includegraphics[trim=5cm 8cm 7cm 9cm, scale=.4]{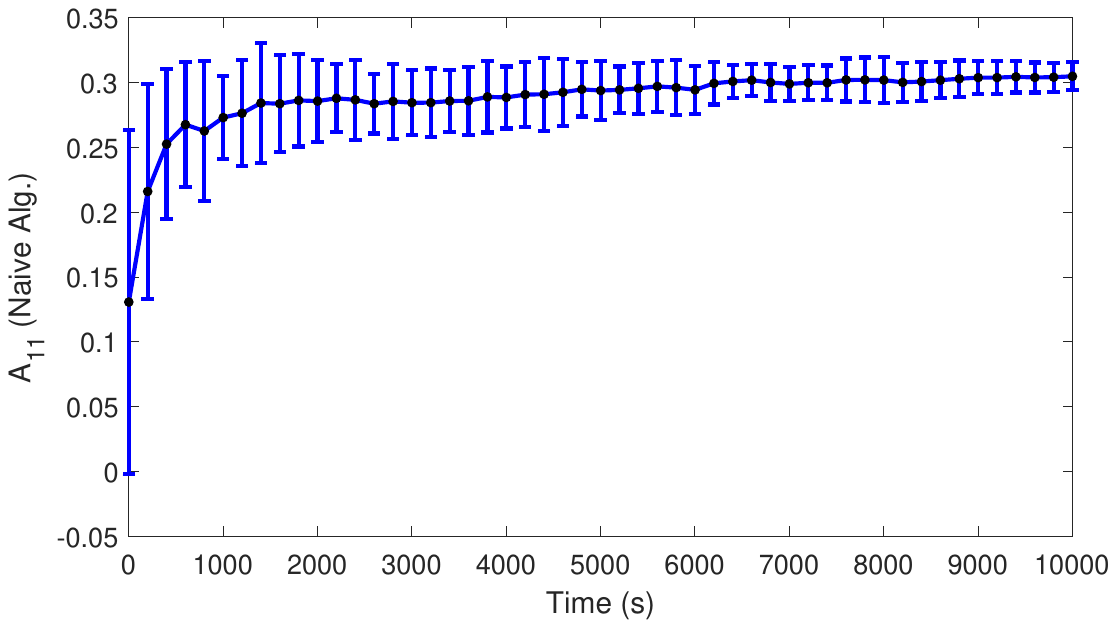}
	\end{subfigure}
	
	\begin{subfigure} 
		\centering
		\hspace{1pt}\includegraphics[trim=5cm 8cm 7cm 9cm, scale=.4]{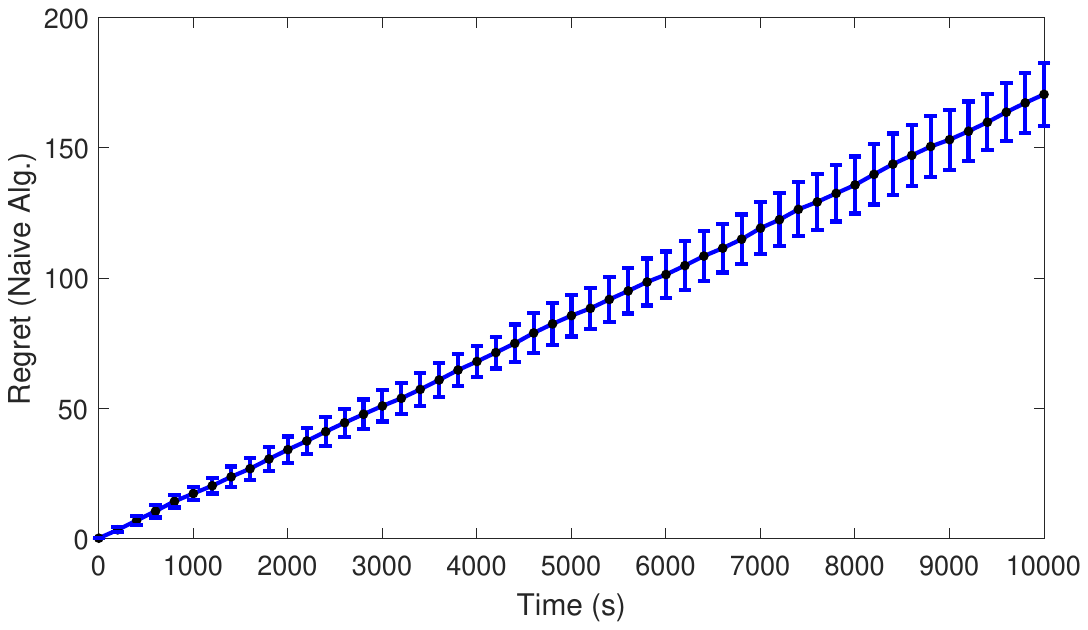}
	\end{subfigure}
	\hfill
	\caption{Performance degradation of Algorithm 1 in the presence of database attack.
Top: $(1,1)$ entry of matrix $A$ corresponding to the estimate $\tilde{\Theta}_t^a$. As can be seen, the estimate converges to an incorrect value with respect to $A_{11}^*$. Bottom: regret as a function of time. Notice the linear dependency.}
	\label{fig:naive_attacked_estimateANDregret}
\end{figure}

An example of this behavior is provided in Figure \ref{fig:naive_attacked_estimateANDregret} which displays the estimate (top) and regret (bottom) of a 'naive' controller designed using Algorithm 1 as-is in the face of data attacks of the kind described above.
For this simulation we consider a control system (adapted from \cite{abbasi2013online}):
\begin{align}
&	x_{t+1}=A_{*} x_{t}+ B_{*} u_t+\omega_{t+1}\\
	\nonumber &	A_{*}=\begin{pmatrix}
		0.54 &-0.11 \\
		-0.026 & 0.63 
	\end{pmatrix},\;\;\;\;\;\;B_{*}=\begin{pmatrix}
		-25 &4.4 \\
		-2.5 & 2.8 
	\end{pmatrix}\times 10^{-4},
		\label{eq:sym_model}
\end{align}
and assume that Assumption 1 holds. For the control cost we choose the cost matrices as follows:
\begin{align*}
	Q=\begin{pmatrix}
		5 &0 \\
		0 & 1 
	\end{pmatrix},\;\;\;\;\;\;R=\begin{pmatrix}
		1/50^2 &0 \\
		0 & 0.1^6 
	\end{pmatrix}.
\end{align*}
The inputs to the OFU algorithm are $T=10000$, $\delta=1/T$, $\lambda=1$, $L=0.1$, $\vartheta=1$ and we repeat simulation $10$ times. Details regarding how to solve the constrained optimization problem that is part of Algorithm 1 are discussed in Section \ref{sec:simulation}.

As can be seen in Figure \ref{fig:naive_attacked_estimateANDregret}, due to the poor performance of the algorithm in parameter estimation the regret of naive algorithm becomes linear under attack, even though, as shown in (\ref{eq:regret_unaatacked}), it is provably sub-linear $O(\sqrt{T})$ in the non-attacked case. For this simulation we observe that the confidence set of naive algorithm fails to capture the true and unknown system parameters for 8377.2 out of 10,000  time steps in average. 

\section{Self-Correcting Algorithm}
\label{section4}

Having observed the degradation of closed-loop performance of Algorithm 1 in the presence of database attacks, our goal, in the remainder of this paper, is to introduce modifications to ensure an appropriate behavior. More precisely, assuming that the attacker performs a database attack $N_{ab}(t)$ times in time interval $[0,\;t]$ 
\begin{enumerate}
    \item the true parameters do lie in the confidence set with high probability i.e., we want to specify $C_t(\delta)$ such that $\Theta_*\in C_t(\delta)$ with high probability.
    \item the regret over horizon $[0,\; T]$ can be a priori bounded in terms of $T$ and $N_{ab}(t)$.
\end{enumerate}

We show that this can be achieved by retaining the general structure outlined in Algorithm 1 while modifying the confidence bound $\beta_t(\delta)$ on the estimate $\hat{\Theta}_t^a$. We call the resulting controller "self-correcting" because it continuously modifies the value of $\beta_t(\delta)$  to account for the effects of database attacks. Since this new bound is used to define the set $C_t(\delta)$ which explicitly appears in the algorithm, it must be computable without precise knowledge of the attack signal and only in terms of the a priori bound $\Lambda$.

Before stating our main results, we give the following lemma, adapted from \cite{abbasi2011online} which gives a self-normalized bound for vector-valued martingales. We will later use this result in our proofs.

\begin{lem} \label{Self_normalized_Bound}
	Let $\mathcal{F}_k$ be a filtration, $\bar{z}_k$ be a vector-valued stochastic process adapted to $\mathcal{F}_k$ and $\omega_k$ be real-valued martingale difference, again adapted to filtration $\mathcal{F}_k$ which satisfies the conditionally sub-Gaussianity assumption (Assumption 2.4) with known constant $L$. 
	Consider the martingale and co-variance matrices:
		\begin{align*}
		S_t:=\sum _{s=1}^{t} \bar{z}_{s-1}\omega_s=\bar{Z}^T_tW_t, \quad \bar{V}_{t}=\lambda I+\sum _{s=1}^{t-1}\bar{z}_{s} \bar{z}_{s}^T
	\end{align*}
	then with probability at least $1-\delta$, $0<\delta<1$ we have,
	\begin{align}
		\left\lVert S_t\right\rVert^2_{\bar{V}_{t}^{-1}} \leq 2 L^2 \bigg(\frac{\det(\bar{V}_t)^{1/2} \det(\lambda I)^{-1/2}}{\delta}\bigg)
	\end{align}	
\end{lem}
\begin{pf}
	Proof has been provided in Theorem 3 and Corollary 1 in \cite{abbasi2011regret}.
\end{pf}
\subsection{Confidence Set in the Presence of Database Attack}\label{AA}
As explained earlier, our first goal is to provide a new confidence bound on the error of parameter estimates in the presence of a database attack. This is the content of the following theorem.
\begin{thm}\label{thm:Conficence_Set_Attacked}
	Let Assumption 1 hold (in particular $\left\lVert \Theta_*\right\rVert \leq \vartheta$) and assume the linear model represented in (\ref{eq:dynam_by_theta}) satisfies Assumption 2 with a known $L$. Let $\hat{\Theta}^a_{t}$ denote the least square estimate and $\bar{V}_{t}=\lambda I+\sum _{s=1}^{t-1}\bar{z}_{s} \bar{z}_{s}^T$ be co-variance matrix, then with probability at least $1-\delta$ we have:
	\begin{align}
		\operatorname{Tr}((\hat{\Theta}^a_{t}-\Theta_*)^TV_{t}(\hat{\Theta}^a_{t}-\Theta_*))
		\leq \beta^a_t(\delta)\label{eq:confidence_set_attacked_1}
	\end{align}
	where
	\begin{align}
		\nonumber\beta^a_t(\delta) & = 
		\quad\Bigg (nL\sqrt{2\log(\frac{\det(\bar{V}_{t})^{1/2}\det(\lambda I)^{-1/2}}{\delta}}+\\
		&\quad \left\lVert \bar{Z}^T_tH_t \right\rVert_{\bar{V}_{t}^{-1}}+(\sqrt{\lambda}+||\bar{Z}^T_tY_t ||_{\bar{V}_{t}^{-1}}) \vartheta\Bigg)^2\label{eq:confidence_set_attacked}  
	\end{align}
	When the attacker can act at every instant of the horizon of interest (i.e., $N_{ab}(t)=t$), this can be upper-bounded as
		\begin{align}
		\nonumber\beta^a_t(\delta)&\leq\Bigg (nL\sqrt{p\log(\frac{p\lambda+2t((1+C^2)X^2_{a,t}+\Lambda^2)}{p\delta \lambda}})+\\
		&\quad 1/\sqrt{\lambda}\Lambda(\vartheta+1) \sqrt{(n+m)t}+\sqrt{\lambda}\vartheta\Bigg)^2 \label{eq:confidence_set_General_att_part2}
	\end{align}
	where $\max_{1 \leq s \leq T} \left\lVert \eta_s\right\rVert \leq \Lambda$ and $\max_{1 \leq s \leq t} \left\lVert x^a_s\right\rVert \leq X_{a,t}$ (i.e., it is the known upper bound on attack signal $\eta$). When the attacker has  limited budget to perform data intrusion the confidence set radius is upper bounded as
	\begin{align}
		\nonumber\beta^a_t(\delta)&\leq\Bigg (nL\sqrt{p\log(\frac{p\lambda+2t((1+C^2)X^2_{a,t}+\Lambda^2)}{p\delta \lambda}})+\\
		&\quad 1/\sqrt{\lambda}\Lambda(\vartheta+1) \sqrt{(n+m)N_{ab}(t)}+\sqrt{\lambda}\vartheta\Bigg)^2. \label{eq:confidence_set_General_att_part2_budg}
	\end{align}
\end{thm}
	\begin{pf}
	Substituting (\ref{eq:dyPttt}) into (\ref{eq:LSE_sol_att}) gives:
	\begin{align}
		\hat{\Theta}^a_{t} &=(\bar{Z}^T_t\bar{Z}_t+\lambda I)^{-1}\bar{Z}^T_t(Z^{a}_t \Theta _{*}+W_t+H_t). \label{eq:obss}
	\end{align}
	And then applying (\ref{eq:BarZi_Z^aRel}) yields, 
	\begin{align}
		\nonumber\hat{\Theta}^a_{t} &=(\bar{Z}^T_t\bar{Z}_t+\lambda I)^{-1}\bar{Z}^T_t((\bar{Z}_t-Y_t) \Theta _{*}+W_t+H_t)\\
		\nonumber&\quad =(\bar{Z}^T_t\bar{Z}_t+\lambda I)^{-1}\bar{Z}^T_t(W_t+H_t)-(\bar{Z}^T_t\bar{Z}_t+\lambda I)^{-1}\bar{Z}^T_t Y_t\Theta_{*}+\\
		&\quad(\bar{Z}^T_t\bar{Z}_t+\lambda I)^{-1}\bar{Z}^T_t\bar{Z}_t\Theta_{*}\label{eq:tpcd}
	\end{align}
	The last term in right hand side can be rewritten as follows:	
	\begin{align*}
		(\bar{Z}^T_t\bar{Z}_t+\lambda I)^{-1}\bar{Z}^T_t\bar{Z}_t\Theta_{*}&=(\bar{Z}^T_t\bar{Z}_t+\lambda I)^{-1}(\bar{Z}^T_t\bar{Z}_t+\lambda I)\Theta_{*}-\\
		&\quad \lambda (\bar{Z}^T_t\bar{Z}_t+\lambda I)^{-1}\Theta_{*} 
	\end{align*}
	now, using $\bar{V}_{t}=\bar{Z}^T_t\bar{Z}_t+\lambda I$
	(\ref{eq:tpcd}) yields:
	\begin{align}
		\hat{\Theta}^a_{t}-\Theta_{*} &=\bar{V}_{t}^{-1}\bar{Z}^T_t(W_t+H_t)+
		\bar{V}_{t}^{-1}(\lambda I+\bar{Z}^T_t Y_t)\Theta_{*}.
	\end{align}
	For an arbitrary random covariate $z$ we have,
	\begin{align}
	\nonumber	z^T\hat{\Theta}^a_{t}-z^T\Theta_{*} &=\langle\,z,\bar{Z}^TW\rangle_{\bar{V}_{t}^{-1}}+\langle\,z,\bar{Z}^TH\rangle_{\bar{V}_{t}^{-1}}\\
		&
		+\langle\,z,(\lambda I+\bar{Z}^T Y)\Theta_{*}\rangle_{\bar{V}_{t}^{-1}}.
	\end{align}
	 By taking norm on both sides one can write,
	\begin{align}
	\nonumber\left\lVert z^T\hat{\Theta}^a_{t}-z^T\Theta_{*}\right\rVert &\leq \left\lVert z\right\rVert_{\bar{V}_{t}^{-1}}\Bigg (\left\lVert \bar{Z}^T_tW_t \right\rVert_{\bar{V}_{t}^{-1}}+\left\lVert \bar{Z}^T_tH_t \right\rVert_{\bar{V}_{t}^{-1}}\\
	\nonumber&+\left\lVert \lambda I+\bar{Z}^T_tY_t \right\rVert _{\bar{V}_{t}^{-1}} \vartheta  \Bigg) \leq \left\lVert z\right\rVert_{\bar{V}_{t}^{-1}}\Bigg (\left\lVert \bar{Z}^T_tW_t \right\rVert_{\bar{V}_{t}^{-1}}\\
	&+\left\lVert \bar{Z}^T_tH_t \right\rVert_{\bar{V}_{t}^{-1}}+(\sqrt{\lambda}+||\bar{Z}^T_tY_t ||_{\bar{V}_{t}^{-1}}) \vartheta\Bigg).
	\end{align}
	Using Lemma 1, $\left\lVert \bar{Z}^T_tW_t \right\rVert_{\bar{V}_{t}^{-1}}$ is bounded from above as:
	\begin{align}
	\left\lVert \bar{Z}^T_tW_t \right\rVert_{\bar{V}_{t}^{-1}} &\leq nL\sqrt{2\log(\frac{\det(\bar{V}_{t})^{1/2}\det(\lambda I)^{-1/2}}{\delta}}).
	\end{align}
	As a result, with probability at least $1-\delta$ we will have,
	\begin{align}
	\nonumber\left\lVert z^T\hat{\Theta}^a_{t}-z^T\Theta_{*}\right\rVert & \leq \left\lVert z\right\rVert_{\bar{V}_{t}^{-1}}
	\Bigg (nL\sqrt{2\log(\frac{\det(\bar{V}_{t})^{1/2}\det(\lambda I)^{-1/2}}{\delta}})+\\
	&\quad \left\lVert \bar{Z}^T_tH_t \right\rVert_{\bar{V}_{t}^{-1}}+(\sqrt{\lambda}+||\bar{Z}^T_tY_t ||_{\bar{V}_{t}^{-1}})\vartheta\Bigg) \label{eq:Conf1}
	\end{align}
	In particular, choosing $z=\bar{V}_{t}(\hat{\Theta}_{t}^a-\Theta_{*})$ and plugging it into (\ref{eq:Conf1}) yields:
	\begin{align}
	\nonumber\left\lVert \hat{\Theta}^a_{t}-\Theta_{*}\right\rVert^2_{\bar{V}_{t}} & \leq \left\lVert \bar{V}_{t}(\hat{\Theta}_{t}^a-\Theta_{*})\right\rVert_{\bar{V}_{t}^{-1}}\\
	\nonumber&\quad\Bigg (nL\sqrt{2\log(\frac{\det(\bar{V}_{t})^{1/2}\det(\lambda I)^{-1/2}}{\delta}})+\\
	&\quad\left\lVert \bar{Z}^T_tH_t \right\rVert_{\bar{V}_{t}^{-1}}+(\sqrt{\lambda}+||\bar{Z}^T_tY_t ||_{\bar{V}_{t}^{-1}})\vartheta\Bigg)
	\end{align}
	and since $ \left\lVert \hat{\Theta}^a_{t}-\Theta_{*}\right\rVert_{\bar{V}_{t}}=\left\lVert \bar{V}_{t}(\hat{\Theta}_{t}^a-\Theta_{*})\right\rVert_{\bar{V}_{t}^{-1}}$, the statement (\ref{eq:confidence_set_attacked_1} \ref{eq:confidence_set_attacked}) holds.Upper-bounding the terms $||\bar{Z}^TH||_{\bar{V}^{-1}}$ and $||\bar{Z}^TY||_{\bar{V}^{-1}}$shows the second statement of theorem, (\ref{eq:confidence_set_General_att_part2}).
  \begin{align*}
||\bar{Z}^TH||^2_{\bar{V}^{-1}}&=\mathrm{Tr}(H^T\bar{Z}\bar{V}^{-1}\bar{Z}^TH)\\
&\quad=\mathrm{Tr}(\bar{Z}\bar{V}^{-1}\bar{Z}^THH^T)\\
&\quad \leq \Lambda^2 t \mathrm{Tr}(\bar{Z}\bar{V}^{-1}\bar{Z}^T)\\
&\quad =\Lambda^2 t \mathrm{Tr}((\lambda I+\bar{Z}^T\bar{Z})^{-1}\bar{Z}^T\bar{Z})
\\
&\quad\leq (n+m) \Lambda^2 t
\end{align*}
	  where in second equality we applied the cyclic property of trace, in the first inequality we applied $\mathrm{Tr}(AB)\leq \mathrm{Tr}(A)\mathrm{Tr}(B)$	which holds for positive definite matrices $A$ and $B$. The last inequality holds for $\lambda>1$.
	Similarly we have the same analysis to the term $||\bar{Z}^TY||^2_{\bar{V}^{-1}}$ which completes the proof. Proof of (\ref{eq:confidence_set_General_att_part2_budg}) follows the same steps with limited attack budget $N_{ab}(t)$ consideration.
	\end{pf}
For the regret bound analysis section, we need a neat upper bound of (\ref{eq:confidence_set_attacked_1}) in terms of $ \Lambda$, $X_{a,t}$ and $t$. Note that the order of time dependency of $\beta^a$ affects the order of regret bound as it will be discussed in the next section.
\subsection{Regret Bound Analysis of Attacked System}\label{BB}
In this part, we analyze the regret bound of the self-correcting controller under database attack. The general form of regret bound analysis for an unattacked system presented in (\ref{eq:regret_unat}-\ref{eq:R3}) is directly applicable to this case as well except for the fact that the attack signal $\eta$ is unknown a priori. This requires us to establish new upper bounds on counterparts of $R_1$, $R_2$, and $R_3$ under attack. However, before proceeding to this, we are required to define the associated "good events" for the attacked system setting. Similar to (\ref {eq:GoodEven_state_unat}) and (\ref{eq:GoodEven_set_unat}) we define:
\begin{subequations}
	\begin{align}
		F^a_{t}=\{\omega \in\Omega  \mid \forall s \leq t, \left\lVert x^a_{s}\right\rVert \leq \alpha^a_{t} \}\label{eq:GoodEven_a_state_unat} \\
		E^a_{t}=\{\omega \in\Omega  \mid \forall s \leq t, \Theta_{*} \in C^a_{s}(\delta/4)  \}\label{eq:GoodEven_a_set_unat} \end{align}
\end{subequations}
where $C^a_{t}$ is defined as follows:
\begin{align}
	C^a_{t}(\delta) =\{\Theta \in \mathbb{R}^{n \times (n+d)}  \mid \operatorname{Tr}((\hat{\Theta}^a_{t}-\Theta)^T\bar {V}_{t}(\hat{\Theta}^a_{t}-\Theta))\leq \beta^a_{t}(\delta)  \} \label{eq:Conf_Set_a_unaata} 
\end{align}
and where $\beta^a_{t}$ is defined by Theorem \ref{thm:Conficence_Set_Attacked} and $\alpha^a_{t}$ is specified in the Appendix. The following lemma bounds ${1}_{(F^a\cap E^a)}R_1$ where $F^a=F^a_T$, $E^a=E^a_T$, and ${1}_A$ is the indicator function of event $A$. 
\begin{lem} \label{lemma_1}
	Let $R_1$ be the counterpart of (\ref{eq:R1}) under attack. Then, with probability at least $1-\delta/2$ we have:
	\begin{align}
		{1}_{(F^a\cap E^a)} R_1\leq 2D W^2 \sqrt{2T \log \frac{8}{\delta}}+n\sqrt{B^\prime_{a,\delta}} 
	\end{align}	
	where $W=Ln\sqrt{2n \log (8nT/\delta)}$ and 
	\begin{align}
		\nonumber B^\prime_{a,\delta}=\big(\nu+TD^2S^2X^2_{a,T}(1+C^2)\big)\times\\
		\log \bigg(\frac{4nv^{-1/2}}{\delta} \sqrt{\nu+TD^2S^2X^2_{a,T}(1+C^2)}\bigg)
	\end{align}	
\end{lem}
\begin{pf}
	Proof follows from lemmas 6 and 7 in \cite{abbasi2011regret}.
\end{pf}
Consider (\ref{eq:R2}) and Algorithm 1, it is clear that most of the terms of $R_2$ take zero value except those times that algorithm has a switch in policy (i.e., where the condition on line 5 of Algorithm 1 is satisfied). The following lemma provides an upper bound for $R_2$ conditioned to the satisfaction of both good events $F^a\cap E^a$.
\begin{lem}\label{thm:lemma2}
	Let $R^a_2$ be the counterpart of (\ref{eq:R2}) under attack. Then, we have:
	\begin{align}
		{1}_{(F^a\cap E^a)}|R_2|\leq  2D X^2_{a,T}(n+d) \log_2 (1+2T/\lambda(X^2_{a,T}(1+C^2)+\Lambda)) \label{eq:boundofR2a}
	\end{align}	
\end{lem}
\begin{pf}
	The proof follows the same steps as lemma 1 in \cite{abbasi2011regret}.
	Let us assume at time steps  $t_{n_1},...,t_{n_N}$ the algorithm 1  changes the policy. Therefore, we have $ \det(V_{t_{n_1}}) \geq 2\lambda^{n+m} $ and $ \det(V_{t_{n_N}})\geq 2^{N-1} \det(V_{t_{n_1}})$. On the other hand we have:
	\begin{align}
		\nonumber\lambda_{max}(V_T)&\leq \lambda +\sum_{s=0}^{T-1} \left\lVert
		\bar{z}_s\right\rVert^2
		\\
		\nonumber&\quad=\lambda + \sum_{s=0}^{T-1} \left\lVert
		z^a_s+\zeta_s\right\rVert^2\\
		\nonumber&\quad \leq \lambda + 2\sum_{s=0}^{T-1}( \left\lVert
		z^a_s\right\rVert^2+\left\lVert
		\zeta_s\right\rVert^2)\\
		&\quad \leq\lambda+2T(X^2_{a,T}(1+C^2)+\Lambda^2_T)  \label{eq:Ineq1} 
	\end{align}
	Then, since $\scriptstyle det(V_T) \leq \lambda_{max}(V_T)^{n+d}$, we deduce that the number $N$ of policy switches is upper bounded as
	\begin{align}
		N \leq (n+m) \log_2 (1+2T/\lambda(X^2_{a,T}(1+C^2)+\Lambda))\label{eq:max_num_switch}
	\end{align}
	Inequality (\ref{eq:boundofR2a}) then results from incorporating the upper bound on 
	$\left\lVert P(\Theta)\right\rVert$. 
\end{pf}
To bound the term $R_3$ we follow same steps as in \cite{abbasi2011regret}. However, we are required to express this bound in terms of states and extended states of the attacked system, $x^a_t$ and $z^a_t$ respectively. We borrow Lemma \ref{thm:lemma 4} from \cite{abbasi2011online}. 
\begin{lem}\label{thm:lemma 4}
	The following holds for $t \geq 1$:
	\begin{align}
		\sum_{s=0}^{t-1}\Bigg(\left\lVert\bar{z}_s\right\rVert^2_{\bar{V}^{-1}_s}\wedge 1 \Bigg) \leq 2 \log \frac{det(\bar{V}_t)}{det(\lambda I)}  
	\end{align}	
	where $(a\wedge b)$ denotes the minimum of $a$ and $b$.
\end{lem}
\begin{pf}
The proof is in \cite{abbasi2011online}.
\end{pf}

To bound the term $|R_3|$, it is required to find an upper bound for the summation $\sum_{s=0}^{T}\left\lVert(\Theta_{*}-\tilde{\Theta}^a_{s})z^{a}_s\right\rVert^2$. The following lemma provides this upper bound.
\begin{lem}\label{thm:Lemma 5}
	The following holds for $t \geq 1$:
	\begin{align}
		\nonumber\sum_{s=0}^{T}\left\lVert(\Theta_{*}-\tilde{\Theta}^a_{s})z^{a}_s\right\rVert^2 &\leq\frac{16}{\lambda}((1+C^2)X^2_{a,T}+\Lambda^2_T)\beta^a_T(\delta/4)\\&\quad\Bigg(2\log \frac{det(\bar{V}_T)}{det(\lambda I)}+
		\sum_{s=0}^{T}\Big(\left\lVert\zeta_s\right\rVert^2_{\bar{V}^{-1}_s}\wedge \frac{1}{2} \Big)\Bigg)
	\end{align}	
\end{lem}
\begin{pf}
	From Lemma 12 of \cite{abbasi2011regret} we can first write 
	\begin{align*}
	\sum_{s=0}^{T}\left\lVert(\Theta_{*}-\tilde{\Theta}^a_{s})z^{a}_s\right\rVert^2 \leq   
		\frac{8}{\lambda}((1+C^2)X^2_{a,T}+\Lambda^2_T)\beta^a_T(\delta/4)\\
		\sum_{s=0}^{T}\Bigg(\left\lVert z^a_s\right\rVert^2_{\bar{V}^{-1}_s}\wedge 1 \Bigg)
		= \frac{8}{\lambda}((1+C^2)X^2_{a,T}+\Lambda^2)\beta^a_T(\delta/4)\\
		\sum_{s=0}^{T}\Big(\left\lVert\bar{z}_s-\zeta_s\right\rVert^2_{\bar{V}^{-1}_s}\wedge 1 \Big)\leq \frac{8}{\lambda}((1+C^2)X^2_{a,T}+\Lambda^2)\beta^a_T(\delta/4)\\
		\sum_{s=0}^{T}\Big(2\left\lVert\bar{z}_s\right\rVert^2_{\bar{V}^{-1}_s}+2\left\lVert\zeta_s\right\rVert^2_{\bar{V}^{-1}_s}\wedge 1 \Big)\leq\frac{8}{\lambda}((1+C^2)X^2_{a,T}+\Lambda^2)\beta^a_T(\delta/4)\\
		\Bigg(2\sum_{s=0}^{T}\Big(\left\lVert\bar{z}_s\right\rVert^2_{\bar{V}^{-1}_s}\wedge \frac{1}{2} \Big)+2\sum_{s=0}^{T}\Big(\left\lVert\zeta_s\right\rVert^2_{\bar{V}^{-1}_s}\wedge \frac{1}{2} \Big)\Bigg)\\
		\leq\frac{8}{\lambda}((1+C^2)X^2_{a,T}+\Lambda^2_T)\beta^a_T(\delta/4)\\\Bigg(2\sum_{s=0}^{T}\Big(\left\lVert\bar{z}_s\right\rVert^2_{\bar{V}^{-1}_s}\wedge 1 \Big)+
		2\sum_{s=0}^{T}\Big(\left\lVert\zeta_t\right\rVert^2_{\bar{V}^{-1}_s}\wedge \frac{1}{2} \Big)\Bigg)
	\end{align*}	
	Substituting the first term of the last expression by the result of Lemma \ref{thm:lemma 4}, completes the proof. 
\end{pf}
The following Lemma provides the upper bound for $|R_3|$.
\begin{lem} \label{thm:Lemma 6}
	Let $R_3$ be the counterpart of (\ref{eq:R3}) under attack, then 
	\begin{align}
		\nonumber {1}_{(F^a\cap E^a)}|R_3|&\leq\frac{8}{\sqrt{\lambda}}((1+C^2)X^2_{a,T}+\Lambda^2)SD\sqrt{\beta^a_T(\delta/4)}\\  &\quad\Bigg(2\log \frac{det(\bar{V}_T)}{det(\lambda I)}+
		\sum_{s=0}^{T}\Big(\left\lVert\zeta_s\right\rVert^2_{\bar{V}^{-1}_s}\wedge \frac{1}{2} \Big)\Bigg)^{1/2}\sqrt{T} \label{eq:boundR3_att}
	\end{align}
\end{lem}
\begin{pf}
	As in the proof of Lemma 13 in \cite{abbasi2011regret}, we first write
	\begin{align}
		\nonumber &{1}_{(F^a\cap E^a)}|R_3|\leq {1}_{(F\cap E)}\Bigg(\sum_{s=0}^{T}\left\lVert P(\tilde{\Theta}_s)(\tilde{\Theta}_s-\Theta_{*})^Tz^{a}_s\right\rVert^2\Bigg)^{1/2}\times\\
		& \Bigg(\sum_{s=0}^{T}\Big(\left\lVert P(\tilde{\Theta}_s)^{1/2}\tilde{\Theta}_s^Tz^{a}_s\right\rVert+\left\lVert
		P(\tilde{\Theta}_s)^{1/2}\Theta_*^Tz^{a}_s\right\rVert\Big)^2\Bigg)^{1/2}
	\end{align}
	then apply Lemma \ref{thm:Lemma 5} and the boundedness assumption on  $P$ and $\tilde{\Theta}_t$ to yield the bound in (\ref{eq:boundR3_att}).
\end{pf}
Combining all three inequalities provides an upper-bound
on the full regret $R_a$ of the self-correcting algorithm under attack, as follows.
\begin{thm}\label{thm:RegretBound}
	The upper-bound for regret of corrective algorithm $R_a$ is given by
	\begin{align}
		\nonumber R_a &\leq 2D W^2 \sqrt{2T \log \frac{8}{\delta}}+n\sqrt{B^\prime_{a,\delta}}\\
		\nonumber &\quad+ 2D X^2_{a,T}(n+m) \log_2 (1+2T/\lambda(X^2_{a,T}(1+C^2)+\Lambda))\\
		\nonumber &\quad + \frac{8}{\sqrt{\lambda}}((1+C^2)X^2_{a,T}+\Lambda^2)sD\sqrt{\beta^a_T(\delta/4)}\\  &\quad\Bigg(2\log \frac{det(\bar{V}_T)}{det(\lambda I)}+
		\sum_{s=0}^{T}\Big(\left\lVert\zeta_s\right\rVert^2_{\bar{V}^{-1}_s}\wedge \frac{1}{2} \Big)\Bigg)^{1/2}\sqrt{T} \label{eq:Attack_Reg}
	\end{align}
\end{thm}	
In addition, Lemma \ref{thm:Lemma 7} below makes it possible to further bound the term $det(\bar{V}_t)$ entirely in terms of $\zeta_t$ and $z^a_t$, which will prove useful to estimate the scaling of the right hand side.	

\begin{lem}\label{thm:Lemma 7}
	\begin{align}
		det (\bar{V}_t) \leq \Bigg(\frac{(n+m)\lambda+2 \sum_{k=1}^{t-1}(\left\lVert z^a_k\right\rVert^2+\left\lVert \zeta_k\right\rVert^2)}{n+m}\Bigg)^{n+m}
	\end{align}
\end{lem}
\begin{pf}
	For a positive definite matrix $M$, we have $det(M)\leq det(\bar{M})$ where $\bar{M}$ is a matrix with zero off-diagonal elements and diagonal elements equal to those of $M$. Hence, for $det(\bar{V}_t)$ we have:
	\begin{align}
		\nonumber det(\bar{V}_t)&\leq \prod_{i=1}^{n+m}\Big(\lambda+\sum_{k=1}^{t-1}(z^a_{ki}+\zeta_{ki})^2\Big)\\ \nonumber&\quad\leq \bigg(\frac{\sum_{i=1}^{n+m}\big(\lambda+\sum_{k=1}^{t-1}(z^a_{ki}+\zeta_{ki})^2\big)}{n+m}\bigg)^{n+m}\\ \nonumber&\quad\leq \bigg(\frac{\sum_{i=1}^{n+m}\big(\lambda+2\sum_{k=1}^{t-1}(z^{a2}_{ki}+\zeta^2_{ki})\big)}{n+m}\bigg)^{n+m}\\
		&\quad=\Bigg(\frac{(n+m)\lambda+2 \sum_{k=1}^{t-1}(\left\lVert z^a_k\right\rVert^2+\left\lVert \zeta_k\right\rVert^2)}{n+m}\Bigg)^{n+m}
	\end{align}	
	In second inequality we applied AM-GM inequality and in the third inequality we apply the property $(a+b)^2\leq 2a^2+2b^2$
\end{pf}
With Theorem \ref{thm:RegretBound} and Lemma \ref{thm:Lemma 7} in hand, we are almost ready to bound the regret in terms of  $X_{a,T}$ and $\Lambda$. The next step is to bound $X_{a,T}$ in terms of problem dependent parameters and time. In order to obtain this bound, we need to provide an explicit value for $\alpha_t^a$, which we left unspecified in (\ref{eq:GoodEven_a_state_unat}). This bound is obtained in Theorem \ref{thm:upperboundx} (See Appendix). However, it cannot be directly used in the regret bound as $\alpha^a_t$ itself depends on $\beta^a_t$ and $Z^a_t$ which in turn are $x^a_t$ dependent. Then a further step is required. Given Theorem \ref{thm:upperboundx}, the following theorem gives the requisite upper bound on $X_{a,T}$.

{\begin{thm}\label{thm:Bound_x_neat_att}
	For any any $t$, $1_{F^a_t}\max_{1\leq s\leq t}||x_s||\leq X_{a,t}$ with
	\begin{align}
     X_{a,t}\leq (D_1+D_2\Lambda \sqrt{N_{ab}(T)}+D_3\sqrt{\log\frac{t}{\delta}})^{n+m+1} \label{eq:BoundedX}
   \end{align}
    where $D_1$, $D_2$, and $D_3$ are appropriate time-independent constants.
\end{thm}
\begin{pf}
	Given upper-bound provided by Theorem \ref{thm:upperboundx}, the proof of (\ref{eq:BoundedX}) follows similar steps in \cite{abbasi2011regret}.	
\end{pf}}

We are now in a position to formulate the main result of this section, relating the order of $R_a$ to the nature of the attacks.

{\begin{thm}\label{thm:regretOrder}
	Let $N_{ab}:=N_{ab}(T)$ be number of attacks in time horizon $[1,\;T]$. Then, $R_a = \mathcal{O}((\Lambda \sqrt{N_{ab}})^p\sqrt{T})$ with $p<2+2(n+m+1)$. In particular, if $N_{ab}=T$ (i.e., if attacks can occur at any point over the horizon), $R_a=\mathcal{O}(\Lambda ^pT^{p+\frac{1}{2}})$. 
\end{thm}

\begin{pf}
	A bound on $R_a$ depending solely on $T$ and the problem constants can be obtained from Theorem \ref{thm:RegretBound} by bounding the last term on the right hand side of (\ref{eq:Attack_Reg}). 
	
	From (\ref{eq:confidence_set_General_att_part2_budg}), it is clear that $\sqrt{\beta^a_T(\delta/4)}$ is $\mathcal{O}(\Lambda\sqrt{N_{ab}})$ for limited attacks budget case.  To bound the second critical term $ 
	(\sum_{s=0}^{T}\Big(\left\lVert\zeta_s\right\rVert^2_{\bar{V}^{-1}_s}\wedge \frac{1}{2} \Big)$, we proceed as follows. The term $||\zeta_s||$ is bounded from above and the covariance matrix $\bar{V}$ is summation of positive semi-definite matrices in time (so that  $\bar{V}_t\succ\bar{V}_{t-1}$). We consider the worst case scenario when the covariance matrix is constant, e.g. $\bar{V}=\lambda I \forall t$ which causes the term be $\mathcal{O}(\Lambda ^2N_{ab})$. However this strict assumption is the case when $Z^{a}_t+Y_t=0,  \forall t$ which is not possible given the definitions of $Y_t$ and $Z^{a}_t$. Therefore, except those initial time steps $t$ that $\left\lVert\zeta_s\right\rVert^2_{\bar{V}^{-1}_s}> \frac{1}{2}$ we generally have sub-linear dependency of the term $ 
	(\sum_{s=0}^{T}\Big(\left\lVert\zeta_s\right\rVert^2_{\bar{V}^{-1}_s}\wedge \frac{1}{2} \Big)$ to $\Lambda^2 N_{ab}$. Given this and applying (\ref{eq:BoundedX}) results in regret order $R_a = \mathcal{O}((\Lambda \sqrt{N_{ab}})^p\sqrt{T})$ with $p<2+2(n+m+1)$.  For a specific case when $N_{ab}=T$ it is straightforward to show that the self-correcting algorithm has the regret of $R_a=\mathcal{O}(\Lambda ^pT^{p+\frac{1}{2}})$. 
	
\end{pf}
\section{Simulations}
\label{sec:simulation}
In this section, we investigate the performance of Algorithm 1 for three settings of non-attacked system, self-corrective (aware) and naive (unaware) attacked systems. This self-correcting algorithm, whose properties have been established above is equipped with an adaptive confidence set adjustment that takes into account the possible attack. This feature distinguishes the so-called self-correcting algorithm from the naive one whose performance has already been depicted and discussed in section \ref{sec:ProbFor} and Figure \ref{fig:naive_attacked_estimateANDregret}. We consider the control system (\ref{eq:sym_model}) to examine self-corrective algorithm and unattacked setting in order to carry out a comparison between the three settings.

It has graphically been shown in \cite{abbasi2013online} that the objective function $J(\Theta)=\mathrm{Tr} (P(\Theta))$ is generally non convex and when it comes to one dimensional system (n, m = 1) it is only convex in drift matrix, $A$. Because of this fact, we decided to solve optimization problem (\ref{eq:nonconvexOpt}) using a projected gradient descent method, with basic step
\begin{align}
	\tilde{\Theta}_{t}\leftarrow PROJ_{C_t(\delta)} \bigg(\tilde{\Theta}_{t}-\gamma \nabla_{\Theta} (\mathrm{Tr} (P(\Theta))) \bigg)
\end{align}
where $\nabla_{\Theta}f$ is the gradient of $f$ with respect to $\Theta$. $C_t(\Theta)$ is the confidence set, $PROJ_g$ is Euclidean projection on $g$ and finally $\gamma$ is the step size. Computation of gradient $\nabla_{\Theta}$ as well as formulation of projection has been explicited in \cite{abbasi2013online}, similar to which we choose the learning rate as follows:
\begin{align*}
	\gamma=\sqrt{\frac{0.001}{\mathrm{Tr}(V_t)}}
\end{align*}
We apply the gradient method for 2000 iterations to solve each OFU optimization problem and apply the projection technique until the projected point lies
inside the confidence ellipsoid. Also, 50 steps of random exploratory actions (from a unit ball) are taken at the beginning before applying the main algorithm.  
For the simulation purpose we assume simple martingale difference type of attack signals. Figure \ref {fig:attacked_corrective} shows the performance of the self-correcting algorithm. {As can be seen in the figure, in the presence of attack, this algorithm is able to keep its estimates close to the unknown parameters of the system.} On the other hand, as demonstrated in Figure \ref{fig:naive_attacked_estimateANDregret} the naive algorithm which sticks to the confidence set (\ref{eq:Conf_Set_unaata}), gradually fails to keep its estimates within a reasonable neighborhood of the true parameters' value. 

\begin{figure}
	\centering
	\vspace{5pt}
	\begin{subfigure}
		\centering
		\includegraphics [trim=5cm 8cm 5cm 9cm, scale=.4]{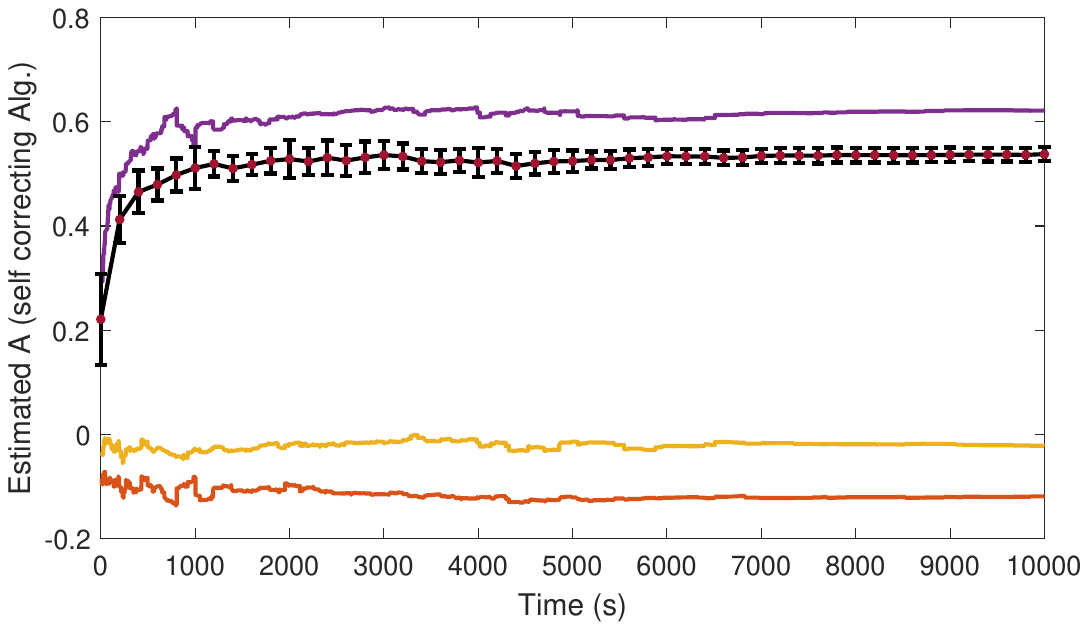}
	\end{subfigure}
	\begin{subfigure}
		\centering
		\includegraphics[trim=5cm 8cm 5cm 9cm, scale=.4]{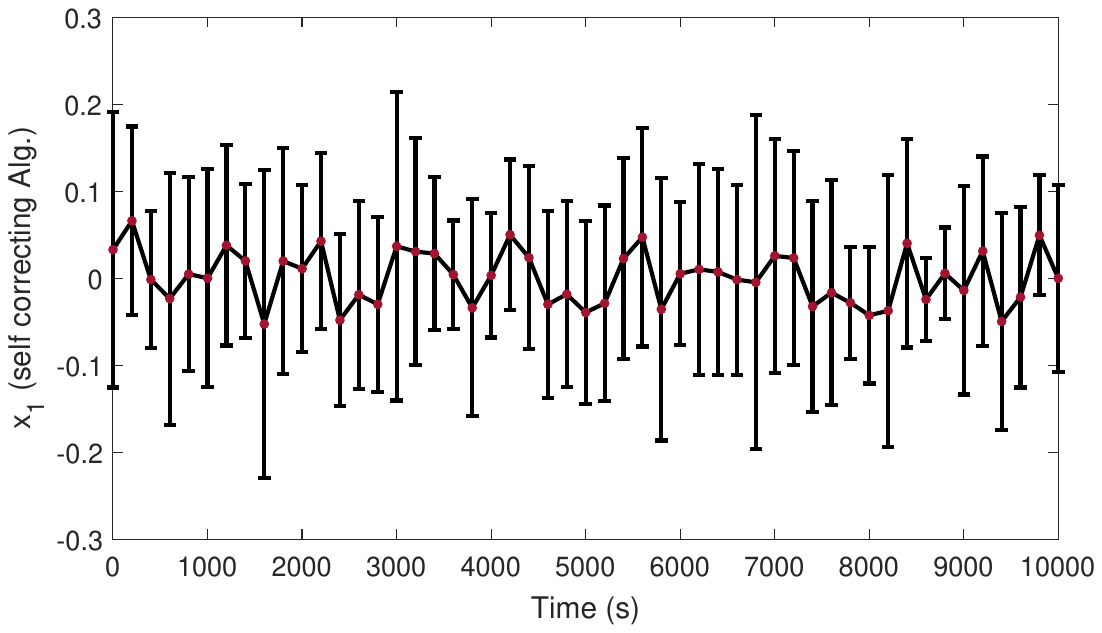}
	\end{subfigure}
	\caption{System parameters' estimates generated by the self-correcting algorithm in the presence of attacks. Top: all entries of estimated parameters $\tilde{A}_t^a$, Bottom: $x_1$ state of the system. Unlike the naive algorithm, the self-correcting algorithm's estimates converge to the correct values of the parameters in the presence of attacks.}
	\label{fig:attacked_corrective}
\end{figure}
 Figure \ref{fig:attacked naive and comparision} presents the regret of the self-correcting algorithm in the absence and presence of attacks. As is expected, the regret for not-attacked case is $\mathcal{O}(\sqrt T)$ and for self-correcting algorithm under attack is sub-linear unlike for the naive algorithm.
 \begin{figure}
 	\centering
 	\vspace{10pt}
 	\begin{subfigure}
 		\centering
 		\includegraphics[trim=5cm 8cm 5cm 9cm, scale=.4]{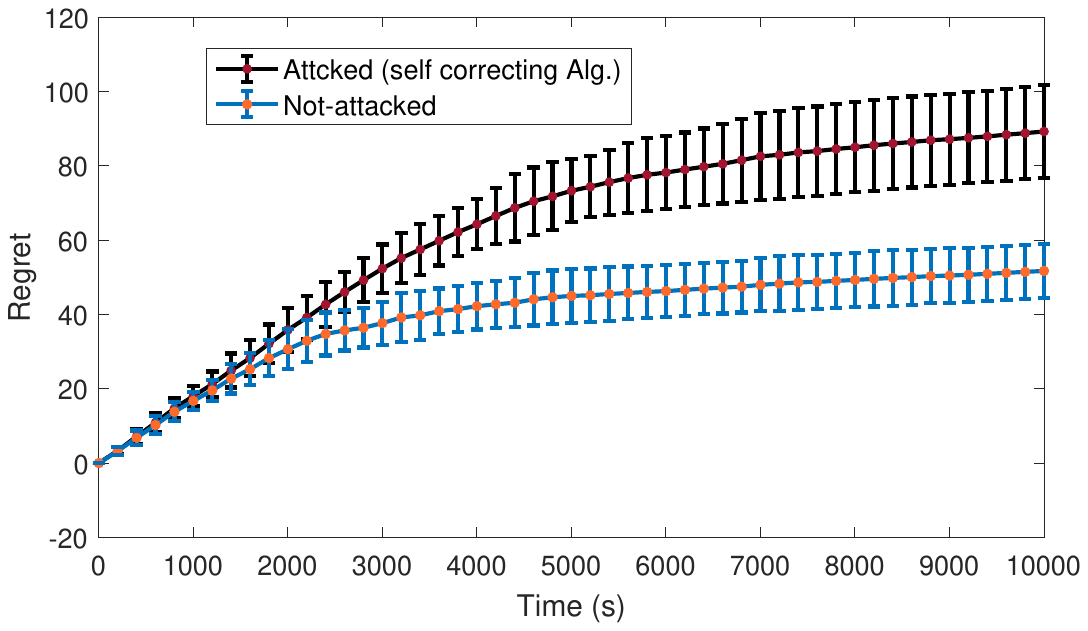}
 	\end{subfigure}
 	\caption{ Regret of the self-correcting algorithm with and without martingale difference attack. }
 	\label{fig:attacked naive and comparision}
 \end{figure}
\section{Conclusion}
In this paper, we have studied database attacks on a class of learning-based LQ controllers and shown that a popular and efficient (in the attack-less setting) algorithm performs poorly when the learning data is modified. We constructed a highly reliable confidence set around unknown parameters that enables the algorithm to keep its estimates close to true parameters in the presence of poisonous data injection into database. Furthermore, a regret bound analysis for the attacked setting is provided to give a measure for attack regret. Simulation results demonstrate the performance of the algorithm equipped with the new confidence set. Our future work will focus on studying the effects of database attacks in conjunction with other simultaneous attacks on sensors and actuators.

\bibliographystyle{plain}        
\bibliography{autosam}           


\newpage
\onecolumn
\section{Appendix}
In this section we aim to obtain $\alpha_t^a$ which provides an upper bound for $x^a_t$.
To start with, following same steps as of the unattacked setting presented in \cite{abbasi2011regret}, our aim is bounding  $\left\lVert
x^a_t\right\rVert$ from above when $\tilde{\Theta}^a_t$ is an estimate of $\Theta_*$ and $E^a$ holds with high probability. In \cite{abbasi2011regret} it has been shown that $\left\lVert (\Theta_*-\tilde{\Theta}_t)^Tz_t\right\rVert$ is well controlled except for a finite number of times. We denote the set of such time instants occurring between $0$ and $T$ as the elements of the set $\tau_T$ which has maximum cardinality of $n+m$. Same justification is applicable for $\left\lVert (\Theta_*-\tilde{\Theta}^a_t)^Tz^a_t\right\rVert$ in the attacked setting. By decoupling the well-controlled and not well-controlled state update rules, the recursion of system can be written as follow:
\begin{align}
	x^a_{t+1}=\Gamma_tx^a_t+r^a_{t+1} \label{eq:up_att}
\end{align}	
where 
\begin{align}
	\Gamma_{t+1}=
	\begin{cases}
		\tilde{A}^a_t+\tilde{B}^a_t K(\tilde{\Theta}^a_t) &\quad t \not\in\tau_T \\
		A_*+B_*K(\tilde{\Theta}^a_t) &\quad t \in \tau_T\\
	\end{cases}
\end{align}	
and 
\begin{align}
	r^a_{t+1}=
	\begin{cases}
		M^a_tz^a_t+\omega_{t+1} &\quad t \not\in\tau_T \\
		\omega_{t+1} &\quad t \in \tau_T\\
	\end{cases} \label{eq:Rs}
\end{align}	
The following theorem provides an upper-bound on the norm $\left\lVert
x^a_t\right\rVert$ which is used to define the event $F^a$.

\begin{thm}\label{thm:upperboundx}
	An upper-bound on the norm of attacked system state, is given by:
\begin{align}
   \nonumber X_{a,t}&\leq  
		\frac{1}{(1-\bar{\alpha})(1-\rho)} \big(\frac{\eta}{\rho}\big)^{n+m}\times\\
	\nonumber	&  \bigg(\tau( (1+C^2)^{\frac{n+m}{2(n+m+1)}}X_{a,t}^{\frac{n+m}{n+m+1}}+\Gamma)+\\
		&(\frac{n+m}{\chi}+2\vartheta)\Lambda_t+L\sqrt{n \log \frac{4nt(t+1)}{\delta}}\bigg):=\alpha_t^a \label{eq:final-2}
\end{align}	
	where $\Gamma=\mathcal{O}(\Lambda \sqrt{N_{ab}(T)})$ and
	\begin{align*}
		Z^a_T =\max\limits_{s\leq T}\left\lVert z^a_s\right\rVert, \quad U=\frac{U_0}{H}, \quad U_0=\frac{1}{16^{n+m-2}(1\vee S^{2(n+m-2)})} 
	\end{align*}
	\begin{align*}
		&  H>\bigg(16\vee\frac{4S^2M\chi^{2(n+m+1)}}{(n+m)U_0}\bigg), \\
		&\tau:= 2\vartheta \sqrt{\frac{n+m}{U}}(U_0H(n+m))^{\frac{n+m}{2(n+m+1)}}
	\end{align*}
		\begin{align*}
    \chi=(n+m)\sqrt{1+C^2}	\frac{1}{1-\rho}\big(\frac{\eta}{\rho}\big)^{n+m}.
        \end{align*}
		\begin{align*}
		&M=\sup_{Y\geq 0}\\ &\frac{\bigg(nL\sqrt{(n+m)\log\big(\frac{1+T(Y+\Lambda)/\lambda}{\delta}\big)}+\frac{\Lambda(s+1)}{\sqrt{\lambda}} \sqrt{(n+m)t}+\sqrt{\lambda}\vartheta\bigg)^2}{Y+\big(\frac{\Lambda(s+1)}{\sqrt{\lambda}} \sqrt{(n+m)t}+\sqrt{\lambda}\vartheta\big)^2}
	\end{align*}
	which allows us to define $\alpha^a_t$ in (\ref{eq:GoodEven_a_state_unat}).	
\end{thm}
To sketch the proof of Theorem \ref{thm:upperboundx} we need Lemmas \ref{thm:Lemma 8} and \ref{lem:uper2}.

\begin{lem}\label{thm:Lemma 8}
	For all $t\leq T$, we have
	\begin{align}
		\max\limits_{0\leq k\leq t}\left\lVert
		M^{a,T}_k\bar{z}_k\right\rVert\leq \beta^a_{t}(\delta/4)^{1/2} 
	\end{align}	
\end{lem}
\begin{pf}
	The proof directly follows from \cite{abbasi2011regret}.		
\end{pf}
The next lemma renders an upper bound for the term   $\max\limits_{0\leq k< t, k\not\in\tau_T}\left\lVert
M^{a,T}_{k}z^a_k\right\rVert$ which is quite similar to bounding $\max\limits_{0\leq k< t, k\not\in\tau_T}\left\lVert
M^T_{k}z_k\right\rVert$ which has been provided in Lemma 18 of \cite{abbasi2011regret}.  

\begin{lem} \label{lem:uper2}
 For all $t\leq T$
	\begin{align}
	\nonumber	\max\limits_{k\leq t, \not\in\tau_t}\left\lVert
		{M^a_{k}}^\top\bar{z}_k\right\rVert&\leq \frac{(n+m)}{\chi}Z^a_t+\frac{(n+m)}{\chi}\Lambda_t\\
    	& +\tau ( {Z_t^a}^{\frac{n+m}{n+m+1}}+\Gamma) \label{eqBound1}
	\end{align}	
	holds true where $\Gamma=\mathcal{O}(\Lambda \sqrt{N_{ab}(T)})$, 

\begin{align*}
    \tau:= 2\vartheta \sqrt{\frac{n+m}{U}}(U_0H(n+m))^{\frac{n+m}{2(n+m+1)}}
\end{align*}
$Z^a_t =\max\limits_{s\leq t}\left\lVert z^a_s\right\rVert$ and
	\begin{align*}
    \chi=(n+m)\sqrt{1+C^2}	\frac{1}{1-\rho}\big(\frac{\eta}{\rho}\big)^{n+m}.
\end{align*}
\end{lem}
\begin{pf}
By following same steps of Lemmas 17 and 18 of \cite{abbasi2011regret} we can write
\begin{align}
    	\nonumber \max\limits_{k\leq t, \not\in\tau_t}\|{M^a_k}^\top \bar{z}_k\|&\leq (n+m)\epsilon_a \bar{Z}_t+\\
    	& 2\vartheta \sqrt{\frac{n+m}{U}}\frac{1}{\epsilon_a ^{n+d}}\beta_t(\delta/4)^{1/2} \label{eq:crucialineq}
\end{align}
where $\bar{Z}_t=\max_{s\leq t}\|\bar{z}_s\|$. We define 
\begin{align*}
  H>\bigg(16\vee\frac{4S^2M\chi^{2(n+m+1)}}{(n+m)U_0}\bigg)
\end{align*}
where 
	\begin{align*}
		&M=\sup_{Y\geq 0}\\ &\frac{\bigg(nL\sqrt{(n+m)\log\big(\frac{1+T(Y+\Lambda)/\lambda}{\delta}\big)}+\frac{\Lambda(s+1)}{\sqrt{\lambda}} \sqrt{(n+m)t}+\sqrt{\lambda}\vartheta\bigg)^2}{Y+\Lambda+\big(\frac{\Lambda(s+1)}{\sqrt{\lambda}} \sqrt{(n+m)t}+\sqrt{\lambda}\vartheta\big)^2}
	\end{align*}
	Now, noting that 
	\begin{align*}
	&\epsilon_a:=\\
	   &\big(\frac{2\vartheta\beta(\delta/4)}{({Z_t+\big(\frac{\Lambda(s+1)}{\sqrt{\lambda}} \sqrt{(n+m)t}+\sqrt{\lambda}\vartheta\big)^2})(n+m)^{\frac{1}{2}}U_0^{\frac{1}{2}}H^{\frac{1}{2}}}\big)^{\frac{1}{(n+m+1)}} \\
	   &\leq\big(\frac{4\vartheta^2M^2}{(n+m)U_0H}\big)^{\frac{1}{2(n+m+1)}}<\frac{1}{\chi}<1
	\end{align*}
	and 

With this definition of $\epsilon_a$,the second term of (\ref{eq:crucialineq}) is written as follows:

\begin{align}
   \nonumber  &2\vartheta \sqrt{\frac{n+m}{U}}\frac{1}{\epsilon_a ^{n+m}}\beta_t(\delta/4)^{1/2} =\\
    \nonumber &2\vartheta \sqrt{\frac{n+m}{U}} \bigg(\frac{(Z^a_t+(\alpha \Lambda \sqrt{t}+\sqrt{\lambda}\vartheta)^{2})U_0^{\frac{1}{2}}H^{\frac{1}{2}} (n+m)^{\frac{1}{2}}}{2\vartheta\beta_t(\delta/4)} \bigg)^{\frac{n+m}{n+m+1}}\\
   \nonumber&\times\beta_t(\delta/4)^{1/2}\leq \tau ( Z_t^{\frac{n+m}{n+m+1}}+\Gamma) \label{eq:secondTerm}
        \end{align}
where $\Gamma=\mathcal{O}(\Lambda \sqrt{N_{ab}(T)})$ and

\begin{align*}
    \tau:= 2\vartheta \sqrt{\frac{n+m}{U}}(U_0H(n+m))^{\frac{n+m}{2(n+m+1)}}
\end{align*}

And for the first term of (\ref{eq:crucialineq}) we simply apply $\bar{Z}_t\leq Z^a_t+\Lambda$. This completes the proof.
\end{pf}

Now, we are ready to sketch the proof of Theorem \ref{thm:upperboundx}.
\begin{pf} (Theorem \ref{thm:upperboundx})
	
	By propagating the state back to time step zero, the state update equation can be written as:
	
\begin{equation}
		x^a_{t}=\prod_{s=0}^{t-1}\Gamma^a_s x^a_0+\sum_{k=1}^{t}\bigg(\prod_{s=k}^{t-1}\Gamma^a_s\bigg) r^a_k\label{eq:Dyn_attp}
	\end{equation}

	From Assumptions 2 and 3 we have; $\scriptstyle \max\limits_{t\leq T}\left\lVert\ \tilde{A}_t+\tilde{B}_t K(\tilde{\Theta}_t)\right\rVert\leq\rho$ and $\scriptstyle\max\limits_{t\leq T}\left\lVert\ A_*+B_*K(\tilde{\Theta}_t)\right\rVert\leq\eta$. Since we have at most $n+m$ not well-controlled system, we can write:
	\begin{align}
		\prod_{s=k}^{t-1}\left\lVert\ \Gamma^a_s\right\rVert&\leq \eta^{n+m}\rho^{t-k-(m+n)}\label{eq:dynBound}
	\end{align}	
	Now, assuming $x^a_0=0$ (without loss of generality) and taking norm from both sides of (\ref{eq:Dyn_attp}) and applying (\ref{eq:dynBound}) we can write:
	\begin{align}
		\nonumber\left\lVert
		x^a_{t}\right\rVert&\leq  
		\frac{1}{1-\rho}\big(\frac{\eta}{\rho}\big)^{n+m} \max\limits_{0\leq k\leq t-1}\big(\left\lVert
		r^a_{k+1}\right\rVert\big) \label{eq:eshgh}
	\end{align}	
	Given (\ref{eq:Rs}), it yields:
	\begin{equation}
		\max\limits_{0\leq k < t}\left\lVert
		r^a_{k+1}\right\rVert\leq \max\limits_{0\leq k< t, k\not\in\tau_T}\left\lVert
		M^{a,T}_{k}z^a_k\right\rVert+
		\max\limits_{0\leq k< t}\left\lVert
		\omega_{k+1}\right\rVert \label{eq:Boundofr}
	\end{equation}	
	
	Finding upper bounds of the terms on right hand side of (\ref{eq:Boundofr}) simply completes the proof. The first term is upper-bounded by using Lemma \ref{lem:uper2} ( \ref{eqBound1}) and as follows
	\begin{align}
	    \nonumber &\max\limits_{0\leq k< t, k\not\in\tau_T}\left\lVert
		M^{a,T}_{k}z^a_k\right\rVert=\max\limits_{0\leq k< t, k\not\in\tau_T}\left\lVert
		M^{a,T}_{k}(z^a_k-\zeta_k)\right\rVert\leq \\
	\nonumber	&\max\limits_{0\leq k< t, k\not\in\tau_T}\left\lVert
		M^{a,T}_{k}z^a_k\right\rVert+\max\limits_{0\leq k< t, k\not\in\tau_T}\left\lVert
		M^{a,T}_{k}\zeta_k\right\rVert\leq \\
		\nonumber&\frac{n+m}{\chi} Z_t^a+\tau({Z_t^a}^{\frac{n+m}{n+m+1}}+\Gamma) +(\frac{n+m}{\chi}+2\vartheta)\Lambda_t\leq\\
		\nonumber &\frac{n+m}{\chi} \sqrt{1+C^2}X_{a,t}+\tau( (1+C^2)^{\frac{n+d}{2(n+d+1)}}X_{a,t}^{\frac{n+m}{n+m+1}}+\Gamma)+\\
		&(\frac{n+m}{\chi}+2\vartheta)\Lambda \label{eq:boundz_bar2}
	\end{align}
	where in the last inequality we applied $Z_t^a\leq \sqrt{1+C^2}X_{a,t}$.
	
	As for the second term, for $1\leq i\leq n$ and $k\leq t$ by sub-gaussianity assumption of $\omega$,  with probability at least $1-\delta/(t(t+1))$ one can write (see \cite{abbasi2011regret}):
	\begin{align*}
		|\omega_{t,i}|\leq L\sqrt{2 \log \frac{t(t+1)}{\delta}}
	\end{align*}
	on some event $G$ and with $P(G)\geq 1-\delta/4$ we have:
	\begin{align}
		\left\lVert
		\omega_t\right\rVert\leq L\sqrt{n \log \frac{4nt(t+1)}{\delta}} \label{eq:boundnoise}
	\end{align}

	Given (\ref{eq:Boundofr}), applying (\ref{eq:boundz_bar2}) and (\ref{eq:boundnoise}) one can write
	
	\begin{align}
	  	\nonumber\left\lVert
		x^a_{t}\right\rVert&\leq  
		\frac{1}{1-\rho}\big(\frac{\eta}{\rho}\big)^{n+m} \bigg(\frac{n+m}{\chi} \sqrt{1+C^2}X_{a,t}+\\
		\nonumber &\tau( (1+C^2)^{\frac{n+m}{2(n+m+1)}}X_{a,t}^{\frac{n+m}{n+m+1}}+\Gamma)+(\frac{n+m}{\chi}+2\vartheta)\Lambda+\\
		&L\sqrt{n \log \frac{4nt(t+1)}{\delta}}\bigg) \label{eq:final-1}
	\end{align}
		Considering the definition of $F^a$, (\ref{eq:final-1}) holds on $G\cap E^a$ and since $G\cap E^a\subset F^a\cap E^a$ holds true, we have $P(G\cap E^a)\leq \delta/2$ and $P(F^a\cap E^a)\leq \delta/2$.
	
	Note that since (\ref{eq:final-1}) holds for $\|x_t^a\|$ $\forall t$, then it holds for $X_{t,a}$ too. Furthermore by definition of $\chi$ we have
	
	\begin{align*}
	    \bar{\alpha}:=\frac{1}{1-\rho}\big(\frac{\eta}{\rho}\big)^{n+m} \frac{n+m}{\chi}\sqrt{1+C^2}<1
	\end{align*}
then we can write	
	
\begin{align}
   \nonumber X_{a,t}&\leq  
		\frac{1}{(1-\bar{\alpha})(1-\rho)} \big(\frac{\eta}{\rho}\big)^{n+m}\times\\
	\nonumber	&  \bigg(\tau( (1+C^2)^{\frac{n+m}{2(n+m+1)}}X_{a,t}^{\frac{n+m}{n+m+1}}+\Gamma)+\\
		&(\frac{n+m}{\chi}+2\vartheta)\Lambda+L\sqrt{n \log \frac{4nt(t+1)}{\delta}}\bigg) \label{eq:final-2}
\end{align}

\end{pf}


\end{document}